# Universal solvent restructuring induced by colloidal nanoparticles

Mirijam Zobel [1*], Reinhard B. Neder [1], Simon A. J. Kimber [2*]

[1] *Lehrstuhl für Kristallographie und Strukturphysik, Friedrich-Alexander University Erlangen-Nürnberg (FAU), Department of Physics, Staudtstr. 3, 91058 Erlangen, Germany.*

[2] *European Synchrotron Radiation Facility, 71 Avenue des Martyrs, 38000 Grenoble, France.*

**Colloidal nanoparticles, used for applications from catalysis and energy applications to cosmetics, are typically embedded in matrixes or dispersed in solutions. The entire particle surface, which is where reactions are expected to occur, is thus exposed. Here we show with x-ray pair distribution function analysis that polar and non-polar solvents universally restructure around nanoparticles. Layers of enhanced order exist with a thickness influenced by the molecule size and up to 2 nanometers beyond the nanoparticle surface. These results show that the enhanced reactivity of solvated nanoparticles includes a contribution from a solvation shell of the size of the particle itself.**

Bulk liquids have long been known to show short range order. The original method of choice was x-ray scattering, which was used by Zachariasen in 1935 to study the short-range order between solvent molecules (*1*). Alcohol molecules were shown to form a hydrogen bonded network within the bulk solvent (*2-4*) and short alkanes were found to align in parallel within domains of ca. 2 nm (*5*). More recently, the influence of hard planar walls on bulk liquids has been investigated (*6-8*). Together with force measurements (*7,8*), x-ray scattering confirmed that an exponentially decaying oscillatory density profile is established near the interface (*9*). With the advent of synchrotron radiation sources, Magnussen et al. showed by x-ray reflectivity that even liquid mercury orders at a flat solid interface in exactly the same way (*9*). More recent examples of such ordering phenomena at interfaces include the restructuring of nonpolar n-hexane (*10*); the assembly of fluorinated ionic liquids at sapphire surfaces (*11*); and the exponentially decaying surface segregation profiles in $Cu_3Au$ alloy interfaces (*12*). These restructuring phenomena, in particular the interlayer spacings and decay lengths, are closely related to the local ordering in the bulk liquid (*9,12*).

Although the reorganization of solvent molecules around isolated cations in solution has also been explored (*13,14*), comparable studies on solvated nanoparticles (NP) are rare, in particular for non-aqueous solvent (*15,16,17*). Solvent molecules are expected and have theoretically been modelled (*15,16*) to rearrange at the liquid-nanoparticle interface, although no definitive experimental proof exists so far. For bulk planar surfaces (*6-8*) and ions (*13,14*) such enhanced order is well understood. Here we report a synchrotron x-ray scattering study of a variety of as-synthesized and commercial NPs in polar and non-polar solvents. We show enhanced ordering of solvent molecules at the NP surface that extends several layers into the bulk liquid. This effect is largely independent of the capping agent, solvent polarity, and particle size.

We systematically redispersed different types of metal and metal oxide nanoparticles in polar and nonpolar organic solvents [see supplementary online materials (SOM) (*18*)]. The influence of the solvent molecule size was studied within the series of primary alcohols (methanol, ethanol, and 1-propanol) and the effect on non-polar solvents with hexane. Fourier transformation of high-energy x-ray scattering patterns yielded the pair distribution functions (PDFs), histograms of all interatomic distances within a sample. Data treatment includes the subtraction of diffraction data of the respective pure solvent so that the signal from the bulk solvent is subtracted prior to the Fourier transformation. The resulting difference PDF (d-PDF) should only contain peaks from intraparticular distances within the NPs and hence be identical to the PDF of a NP powder sample. We observed, however, a distinct, exponentially damped sinusoidal oscillation, which vanished in dry samples. [Fig.1; further examples are shown in fig. S7 and S8 (*18*)].

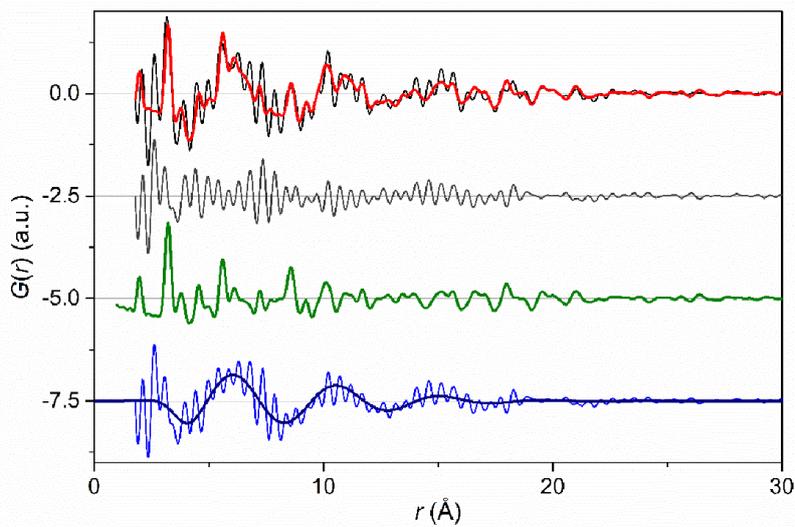

**Figure 1:** Fit to the PDF of redispersed ZnO NPs with citrate ligands in propanol. Experimental d-PDF of ZnO NPs (black) and their fit (red), showing the overall difference of the fit (grey), the contribution of the nanoparticle (green) and the contribution and fit of the restructured solvent (blue) in offset for means of clarity. The contribution of the restructured solvent (blue) is the dd-PDF of the experimental, background corrected d-PDF (black) and the nanoparticle (green).

A least-squares fit (red) of the linear combination of the respective NP powder (green) and an exponentially decaying sinusoidal oscillation (blue) describes all data sets (black) adequately [see SOM (*18*)] and gives access to four parameters that describe the restructuring: i) the amplitude, ii) the position of the first maximum of the oscillation, iii) the wavelength and iv) the modulation depth (for calculation see Materials and Methods). The amplitude contains information about how many solvent molecules restructure and the modulation depth describes how far this enhanced ordering extents into the bulk liquid. The position and wavelength describe the arrangement and layer spacing of the molecules.

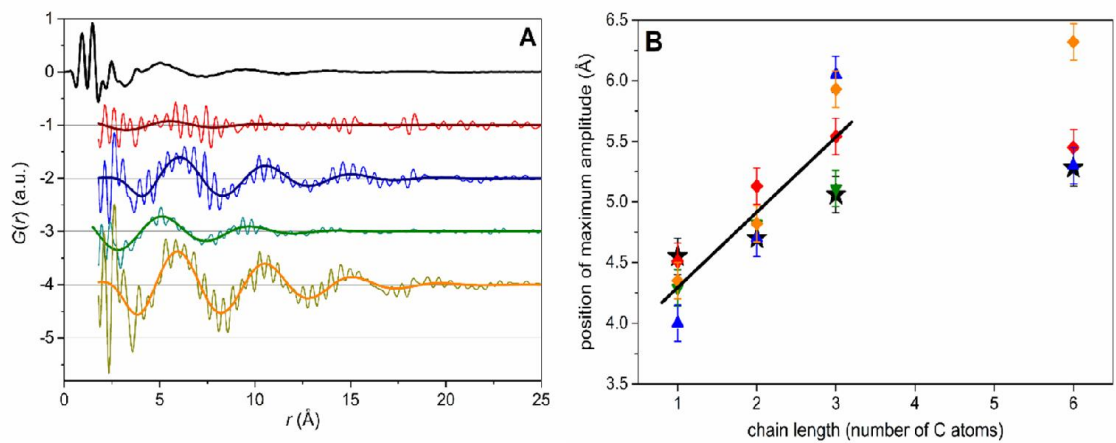

**Figure 2:** Comparison of short range order in restructured and bulk solvents. PDF of pure bulk propanol (black) and dd-PDFs of restructured propanol around ZnO NPs with ligand acetate (red), citrate (blue), dmlt (green) and pent (yellow) (A). The amplitudes of the dd-PDFs are scaled with respect to the contribution of ZnO NPs to the data; Maxima positions of dd-PDFs of the redispersed ZnO NPs in all solvents plotted over the chain length of the solvent molecules (B). The positions follow the trend of the pure solvents (black stars), yet the dd-PDF positions are distinctly offset with respect to the error bars that resulted from least-squares fits.

Fig. 2A shows the PDF of bulk propanol together with the double-difference PDFs (dd-PDFs) of the restructured solvent and their fits for four different ZnO NPs in propanol. Here the dd-PDF is the difference between the d-PDF of the redispersed sample and the PDF of the NP powder. The resulting dd-PDFs differ from the PDF of the pure solvent in that they exhibit a distinctively larger modulation depth. In contrast to bulk propanol, the dd-PDFs do not show sharp intramolecular distances of the propanol molecules in the range 0 to 3 Å, because they are entirely corrected for in the background subtraction. The restructuring does not change the internal molecular structure but only changes the relative orientation between different solvent molecules. Thus, the enhanced short-range order around the NPs becomes visible as the oscillation in the dd-PDF. We observed the same kind of restructuring for a wide range of metal oxide and metal nanoparticles (fig. S5 and S6 (*18*)).

In essence, the PDF represents the distribution of all interatomic distances *r*. As we work in a highly diluted system, we can expect that the structure of the bulk solvent does not change. Likewise, the contribution of the NPs to the PDF does not differ from that of the PDF of the pure NP. As all changes to the experimental dd-PDF occur upon redispersing the NP in the pure solvent, these changes must be induced by the NP surfaces (see SOM for further details (*18*)). Thus, our dd-PDFs represent a change of the short-range order within the solvent near the NP surface. With increasing distance, the signal in the dd-PDF exponentially decays. The orientational averaging intrinsic to the PDF technique prevents us from determining the direction of rearrangement, however, the signal bears a striking similarity to those seen using specular x-ray surface scattering, which probes only the direction normal to the interface.

A comparison of the maximum position, wavelength, and modulation depth of the oscillation for our ZnO NPs is shown in Fig. 2B and the figure S5. Remarkably, a linear trend in the position of the maximum amplitude is found with alcohol chain length. Furthermore, this trend occurs for four different capping agents, showing that surface composition plays a minor role. A larger spread in these parameters is found for other types of nanoparticles (Ag, $ZrO_2$, $TiO_2$ and $In_2O_3$, figure S6). These results show that, to first approximation, and in agreement with MD simulations (*15*), the microscopic details of the nanoparticle surface only weakly influence the solvent restructuring. However, it is possible that the spread of data points in Fig. 2B, which increases for increasing alkyl chain length, indicates an additional role for particle shape, surface restructuring or differences in capping agent.

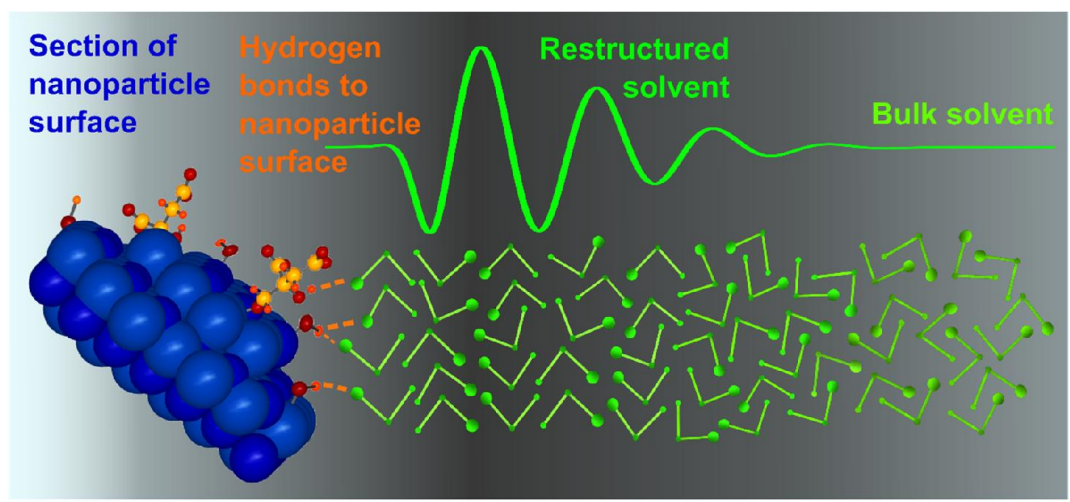

**Figure 3:** Enhanced short range order of solvent molecules at ZnO NP surfaces. The ethanol molecules (hydrogen atoms omitted) form hydrogen bonds with surface hydroxyl groups and citrate molecules. The surface coverage of these groups is reduced for means of clarity. The enhanced short range order extends a few molecular layers into the bulk liquid before bulk properties are recovered. In bulk ethanol, molecules do not arrange in pairs, but form winding-chains or hexamers (*2,3,4*) and the enhanced short range order around the nanoparticles is not as crystalline as suggested in the scheme, which serves as simplified illustrative representation.

Fig. 3 illustrates our model of restructuring for ethanol at the surface of a ZnO NP decorated with citrate and hydroxyl groups of capping ligands. The surface coverage of NPs with organic ligand molecules is sufficient to prevent agglomeration, but in fact is astonishingly small according to neutron PDF data (*19*) and NMR studies (*20*). MD simulations (*15,21,22*), in agreement with experimental evidence from NMR (*20*) suggest that the vast majority of ZnO surface sites are terminated by hydroxyl groups. These could form a hydrogen bonded network with adjacent solvent molecules.

Because of the shift in the oscillation with the solvent size, we conclude that the alcohol molecules tend to align perpendicular to the NP surface. Their hydroxyl groups form hydrogen bonds with the ligand molecules and hydroxyl groups. The alkyl chains of the solvent point away from the NP surface. The next- and second next-neighbouring molecules align such that hydrogen bonds can be formed within the solvent, which results in alternating layers of methyl groups and hydroxyl groups thereby building layers of decreased and enhanced electron density as depicted in Fig. 3. Adjacent molecules within such a layer would orient in parallel as observed for liquid films (*23*). The extent of restructuring depends on the solvent size and packing ability. Here the packing ability is comparable for our alcohols, because the hydroxyl group is always in a terminal position and the alkane chain is not branched. A second-harmonic generation (SHG) study on the interaction of organic solvent and solute molecules with hydroxylated silica surfaces supports our hydrogen bonding model (*24*). This study also showed that nonpolar solvents rearrange at hydroxylated surfaces, which supports our observation that the nonpolar n-hexane restructures at the NP surfaces. However SHG is only sensitive to broken symmetry at interfaces, whereas it cannot provide information on the decay of the restructuring into the bulk liquid as evidenced by our dd-PDFs.

**Acknowledgements**: We want to acknowledge the BMBF under grant no. 05K10WEB and a scholarship of the Friedrich-Alexander-University Erlangen-Nürnberg for financial support. Beam time at the European Synchrotron Radiation Facility and Argonne National Laboratory is gratefully acknowledged. We thank J. Hudspeth from beamline ID-15-B, ESRF as well as K. Chapman and K. Beyer from beamline 11-ID-B, APS for support during our beamtime. We acknowledge A. Magerl and H. Reichert for discussions.
All experimental raw data is stored at the European Synchrotron Radiation Facility. For access, please contact S. A. J. Kimber (kimber@esrf.fr).

**Supplementary Materials**

Materials and Methods
Supplementary Text
Figures S1 – S11
Table S1
References (*25-33*)

# Supplementary Materials

**Materials and Methods**

<u>Synthesis of ZnO NPs.</u> The synthesis procedure is based upon Spanhel and Anderson (*24*) with changes as suggested by (*26-28*). 0.59 g zinc acetate dihydrate (> 99 %) was dissolved in 100 mL anhydrous ethanol (99.5 %, < 0,005 % $H_2O$) to yield a 27 mM solution. Together with the zinc acetate, either none or one additional organic ligand was dissolved at 6.75 mM and stirred for about an hour at room temperature. The organic ligands used are known to influence the shape and size of the final NPs due to selective adsorption on certain crystal faces. These are 1,5-diphenyl-1,3,5-pentanetrione (pent), dimethyl-L-tartrate (dmlt), citrate and acetate (*29,30*). The addition of the organic base tetramethylammonium hydroxide (TMAH, 25 wt% in methanol) at a ratio $[OH^-]/[Zn^{2+}]$ = 2.4 initiates the formation of NPs. Solvents were obtained from either Roth or Sigma Aldrich. Zinc acetate dihydrate, 1,5-diphenyl-1,3,5-pentanetrione and dimethyl-L-tartrate were purchased from Sigma Aldrich and citric acid bought from Roth. All chemicals were used without further purification. The solutions were stirred depending on the nucleation kinetics of the ligand molecules used, i.e. in between 1 to 16 hours. The subsequent addition of a mixture of hexane and acetone (ratio 2.5 : 1) precipitates the NPs. The agglomerated NPs were centrifuged and the resulting gel was dried overnight in a desiccator to gain the nanoparticle powder.

<u>Commercial NPs.</u> NPs have been bought from two different companies with the following specifications: Titanium dioxide particles of 5 nm diameter from nanoAmor (no surface stabilization by organic ligands, anatase structure, Nanostructured & Amorphous Materials, Inc., Houston, TX 77084, USA). 7 nm silver NPs stabilized with alkanethioles, 4 nm indium oxide NPs stabilized with nitrate ions, and 3 nm zirconia dioxide stabilized with benzoic acid were purchased from Plasmachem (PlasmaChem GmbH, Rudower Chaussee 29, D-12489 Berlin, Germany).

<u>Redispersion experiments.</u> The NPs were redispersed with equivalent concentrations in four different solvents: methanol, ethanol, 1-propanol and n-hexane.

<u>Cleaning of glassware.</u> All glassware, which was used for the experiments, was cleaned with 1 wt% NaOH and HCl and rinsed three times with water and Millipore water and dried at 110 °C in a heating oven prior to usage.

<u>High-energy x-ray diffraction.</u> XRD measurements were primarily carried out at beamline ID-15-B at ESRF, Grenoble, with a x-ray energy of 55 keV, using a Perkin Elmer disordered silicon detector and partially at beamline 11-ID-B at APS, Chicago with an equivalent setup. Each nanoparticle was measured as a powder sample in polyimide capillaries ($d_i$ = 1 mm) for reference of the particle size and crystallinity. The redispersed NPs were stirred with a magnetic stirrer for the duration of the experiment to keep the particles suspended. The dispersion was pumped with a peristaltic pump through a stainless steel flow cell with polyimide windows and an x-ray path length of 3 mm. The volume flow was 190 mL/min. The flow cell was cleaned with HCl and Millipore water and dried in a heating oven in between two experiments. The entire tubing was exchanged for each experiment. The experiments were carried out at room temperature and multiframe data sets were typically collected for 10 minutes and averaged. Distance calibration was done with NIST $CeO_2$ standards. Background measurements of the pure solvents were done for 10 minutes while pumping the solvent to guarantee equivalent conditions.

Modelling of PDF data. Data reduction was done with Fit2D (*31*), PDF calculation with PDFgetX3 (*32*). Fourier Transformation of the diffraction data results in the pair distribution function (PDF), $G(r)$.

$$G(r) = \frac{2}{\pi} \int_{Q_{min}}^{Q_{max}} Q[S(Q)-1]\sin(Qr)dQ \qquad (\text{M-1})$$

$Q = 4\pi/\lambda \sin(\theta)$ represents the wave vector transfer with the wavelength $\lambda$ and the scattering angle $2\theta$ and the Fourier transformation runs from the smallest experimentally accessible $Q_{min} = 0.6$ Å$^{-1}$ to a $Q_{max}$, suitably chosen to reduce termination errors. $Q_{max}$ was in between 12.5 and 14 Å$^{-1}$ for the dispersions data and 22.5 Å$^{-1}$ for the powder data. The total scattering structure function $S(Q)$ corresponds to the regular diffraction data after correction for background, experimental artefacts and proper normalization with the scattering cross-sections of all contributing atoms. The chosen $Q_{min} = 0.6$ Å$^{-1}$ guarantees that all essential peaks of each of the pure solvents are within the diffraction pattern.

PDF modelling was done with the DISCUS software package (*33*). ZnO NP powders were modelled as monodisperse ellipsoids of wurtzite structure with variable structural parameters (lattice parameters, etc.) and diameters, omitting the organic ligand molecules. Results of these fits are shown in Table S1. The background corrected PDFs of the redispersed NPs consist of the PDF of the nanoparticle as well as a dampened sinusoidal oscillation. In order to fit the nanoparticle contribution, the experimental powder PDF of the respective nanoparticle was scaled to the PDF of the redispersed sample together with an exponentially dampened sine wave of the form

$$w(r) = A\sin\left(2\pi\left(\frac{r}{\lambda} - \varphi\right)\right) e^{-\left(\frac{(r-r_0)}{2\sigma_{eff}}\right)^2} \qquad (\text{M-2})$$

where $A$ is the amplitude of the oscillation, $\lambda$ the wavelength, $\varphi$ the phase shift and $\sigma_{eff}$ the effective damping with $\sigma_{eff} = \sigma/a$ for $r < r_0$ and $\sigma_{eff} = \sigma*a$ for $r > r_0$, where $a$ is the asymmetry parameter.

This sum was refined with a least squares refinement as well as an evolutionary algorithm as consistency check to all data sets. $r_0$ does not correspond to a physical parameter in real space, but is used for the description of the different damping behaviour of the fitted oscillation for low and high $r$.

The raw data justifies the modelling of the oscillation as exponentially dampened. The background corrected data prior to Fourier transformation, called reduced total scattering structure function $F(Q)$, contains additional peaks, which are not from the NPs and which can be attributed to restructured solvent (fig. S1 and S2). These peaks in $F(Q) = Q[S(Q)-1]$ can be fitted by Pseudo-Voigt functions and hence, the partial Lorentzian character of those peaks demands an exponentially decaying function as implemented above (*16*).

The modulation depth $\sigma$ is calculated from the peak height ratio of the second peak to the first peak of the dampened oscillation as follows:

$$\sigma^2 = \frac{1}{2} \frac{\left((r_1 - r_0)^2 - (r_2 - r_0)^2\right)}{ln\left(\frac{A_2}{A_1}\right)} \quad \text{(M-3)}$$

with $r_1$ and $r_2$ as the positions of the first and second maxima of the oscillations, $r_0$ as defined in equation M-2, $A_1$ and $A_2$ as the amplitudes of the first and second maxima of the oscillations.

The amplitude of the oscillation corresponds to the amount of solvent molecules that are preferentially ordered at the particle surface. Assuming that the NPs redisperse to the same extent, the d-PDFs in Fig. 2a were scaled such that the respective contributions of the NPs, which are not plotted for means of clarity, would be of identical amplitude.

**Confirmation of proper attribution of the dampened sinusoidal oscillation to restructuring effect**

As stated by Magnussen et al. the preferential ordering at interfaces is an enhancement of bulk properties (*13*). Therefore it might be easy to mistake signal from bulk solvent as signal from restructured solvent. However, within these experiments and data analysis, multiple consistency checks have been done to unambiguously interpret the restructuring as such.

1)  No influence of polyimide foil windows of flow cell

There is no influence of the polyimide windows of the flow cell because identical windows are used for the measurements of the solvent background as well as the redispersed particles. Even if the windows induce an addition tiny restructuring at their surface, this restructuring would cancel out by subtracting the solvent background measured under identical conditions.

2)  No influence of x-ray beam

The series of ZnO NPs with ligand dmlt in methanol, ethanol and 1-propanol was measured at 90 keV instead of 55 keV to rule out an influence of the x-ray energy onto the oscillation in the data. Moreover, all data collected during the measuring time of 10 minutes showed identical XRD curves, so that there is no x-ray induced effect or change in the data.

3)  No beamline-specific effect

The restructuring was also observed for ZnO NPs with ligand acetate at beamline 11-ID-B, APS, in order to rule out beamline specific artefacts.

4) The PDFs of the nanoparticle powder samples do not show any oscillation comparable to the ones observed in our dd-PDFs.

Fig. S8 displays the fit (red) of monodisperse wurzitic ZnO nanoparticles with ellipsoidal shape to the experimental powder PDFs (black) of a ZnO NP sample with the ligand citrate (panel A) and ligand pent (panel B). The fit is based on a model of the crystalline NP and did not contain contributions from the ligand. The results of the structural refinement are given in Table S1. All peaks of the experimental PDFs are well described and no modulation similar to any solvent restructuring is observed in the difference curve (black, offset).

5) Restructuring induced by presence of NPs in solvent

The ZnO NPs with ligand acetate were directly redispersed into the pure solvent which was already pumped through the flow cell for the background measurement. No exchange of glassware, tubing or flow cell was performed. Therefore whatever difference occurs in the PDF has to be attributed to the dispersion of the NPs. As the nanoparticle powders were measured separately as well and since none of these powder PDFs features any sinusoidal oscillation it is conclusive that the additional oscillation in the PDF of the redispersed NPs must be induced by the dispersion process, hence the restructuring of the solvent molecules at the nanoparticle surface.

6) Restructuring already visible in a peak shift and FWHM change of the primary peak of the solvents in the XRD patterns (fig. S1).

7) No data analysis error
   - Manual subtraction of the $I(Q)$ data of the bulk solvent from the one of the redispersed NPs results in difference curves similar to the $F(Q)$s produced in PDFgetX3.
   - Manual Fourier transformation of the $F(Q)$ datasets calculated by PDFgetX3 also results in PDFs that contain the same oscillation.
   - Some PDFs were additionally and independently calculated with the PDFgetX2 software and refined separately, resulting in identical oscillations and fit parameters.

8) $Q_{min}$ = 2 Å$^{-1}$ reduces oscillation distinctly
   - We used $Q_{min}$ = 0.6 Å$^{-1}$ in the data reduction, because there is a peak in the background corrected $F(Q)$s in between 0.6 and 2 Å$^{-1}$ (Fig. S2). This peak can directly be correlated to a shift of the peak position and the FWHM of the main solvent peak in the XRD data (Fig. S1). During the background correction, the $I(Q)$ of the pure solvent is subtracted from the $I(Q)$ of the redispersed samples. The initial slight peak change in $I(Q)$ results in a peak in the $F(Q)$, which is quite different in position than the corresponding peak in the $F(Q)$ of the pure solvent (fig. S2).
   - When cutting off the $F(Q)$ data at 2 Å$^{-1}$, the major signal from the restructured solvent is cut off, which reduces the oscillation in the PDF strongly. However, the oscillation is still slightly present, since the restructuring also changes the $F(Q)$ at $Q > 2$ Å$^{-1}$.
- This main peak of the (restructured) solvent in the $F(Q)$ can be integrated using a Pseudo-Voigt function:

$$P = t_0 + t_1 * (x - x_0) + I * (\eta L + (1 - \eta) G) * a \tag{S-1}$$

$$L = \frac{2}{\pi} \frac{\beta}{\left(\beta^2 + 4*(x-x_0)^2\right)} \tag{S-2}$$

$$G = 2 \frac{\sqrt{\frac{\ln(2)}{\pi}}}{\beta} * \exp\left(-4\ln(2) * \frac{(x-x_0)^2}{\beta^2}\right) \tag{S-3}$$

$$a = 1 + \frac{2*a_1*x*\exp(-x^2)}{\tan(x)} \tag{S-4}$$

- $\beta$ is the full width half maximum (FWHM) of the peak, $a$ the peak asymmetry, $P$ the Pseudo-Voigt function, $L$ the Lorentz function and $G$ the Gaussian function.
  In Fig. S4, the resulting peak positions are plotted versus the positions of the first maxima of the oscillations in the PDFs of the redispersed samples. These plots show a nice trend for all solvents. The data points correspond to both ZnO NPs with different ligands as well as Ag, TiO$_2$ and ZrO$_2$. This trend is expected as $G(r)$ is the sine Fourier transform of $F(Q)$. As the first peaks of In$_2$O$_3$ overlap with the solvent peak in $F(Q)$, this analysis could not be performed.

9) No restructuring due to continuous flow and pumping
   - There might be doubt whether the continuous flow of the solvent and redispersed NPs through the flow cell induces ordering. Therefore, all measurements were done with the same pumping speed of the peristaltic pump, which was fixed to 90 rpm, equivalent to a flow rate of 190 mL/min.
   - In order to rule out the possibility that the continuous flow itself leads to restructuring, redispersed ZnO NPs with the ligand pent in propanol were pumped with different speeds of 50, 90 and 150 rpm, e.g. 100, 190 and 300 mL/min. The background was only measured at 190 mL/min. PDFs were calculated for 100, 190 and 300 mL/min pumping speed of the sample with subtraction of the 190 mL/min propanol background. The fit results of the oscillation are shown in fig. S3. Within this histogram plot, these results are compared to the fits results of ZnO NPs with ligand pent in methanol and ethanol. The modulation depth is, in general, less reproducible in the fits in comparison with the reproducibility of the maximum position and the wavelength. The plot clearly shows that the solvent restructuring is not an artefact of the pumping speed since the fit results of the 100, 190 and 300 ml/min propanol measurements show identical values for the maximum and wavelengths, in particular when compared to the corresponding values of methanol and ethanol, both measured at 190 mL/min.

10) Restructuring effect cannot be explained by a pure effect of ligand molecules
   - According to the manufactuer's specification, $TiO_2$ NPs do not have surface ligands at all.
   - The surface coverage of the ligand molecules on the ZnO NP surface is astonishingly small, as theoretically modelled by MD simulations (see refs (*15,21,22*)) and experimentally confirmed by neutron PDF data (*19*) and NMR studies (*20*). We carried out TGA measurements on the ZnO-pent powders and the mass loss, which can be attributed to the disintegration of bound ligand molecules, corresponds to surface coverages as stated in ref. (*20*). Those low surface coverages of ca. 1.7 $nm^{-2}$ for our ligand pent, correspond to ca. 40 pent molecules on the surface of one NP. A solvation shell of e.g. propanol, which extends ca. 2 nm into the bulk, corresponds to 1090 propanol molecules. This huge difference in the amount of scatterers explains why the enhanced order of organic solvent molecules can be observed although the individual scattering power of one molecule does not exceed the scattering power of a single ligand molecule.
   - We calculated x-ray and neutron PDF data for ZnO NP with and without the ligand citrate, see Fig. S9. The surface coverage corresponds to 1.1 citrate molecules per $nm^2$. The modelled ZnO NP with ligand molecules is shown in the inset of panel A. The difference curve of the x-ray PDFs (panel A) of the ZnO NP with (red) and without (black) ligands are shown in offset (black). This difference curve is almost flat for x-rays, because the few ligand molecules are almost invisible for the x-rays. The corresponding plot in panel B shows the neutron PDFs of the identical nanoparticle with (red) and without (black) ligands together with their difference (black, in offset). The neutron difference curve features peaks below 5 Å, which can be attributed to distinct intramolecular distances within the citrate molecule, see inset in panel B. However, no oscillation, which could be mistaken for any solvent restructuring, is observable in either difference, x-ray or neutron. Hence, the bound ligand molecules do not induce any restructuring signal as observed in our dd-PDFs or the redispersed NPs.
   - In order to doublecheck, whether solvent molecules restructure around the ligand molecules at all, we dissolved dmlt and pent molecules in ethanol at a concentration as it was employed during the ZnO NP synthesis. We background corrected for pure solvent and the resulting d-PDFs only feature peaks corresponding to dmlt / pent intramolecular distances, with no modulation at longer distances which would indicate restructuring, see fig. S10. Therefore, no restructuring occurs around redispersed ligand molecules and there is no reason to assume, that ligand molecules on NP behave differently.
- Combining these information, the ligand molecules do not induce solvent restructuring as observed in the dd-PDFs of redispersed NPs.

**Additional Author notes:**
**Author Contributions:** M. Zobel had the idea and designed the experiment. M. Zobel, R.B. Neder and S.A.J. Kimber performed the PDF measurements. M. Zobel analysed the PDF data with help from R.B. Neder and S.A.J. Kimber. M. Zobel discussed the results and wrote the paper, with help from R.B. Neder and S.A.J. Kimber.

**Author Information:** The authors declare no competing financial interests. Readers are welcome to comment on the online version of the paper. Correspondence and requests for materials should be addressed to M. Zobel, mirijam.zobel@fau.de or S.A.J. Kimber, kimber@esrf.fr.

**Figures of Supplementary Materials S1 – S11**

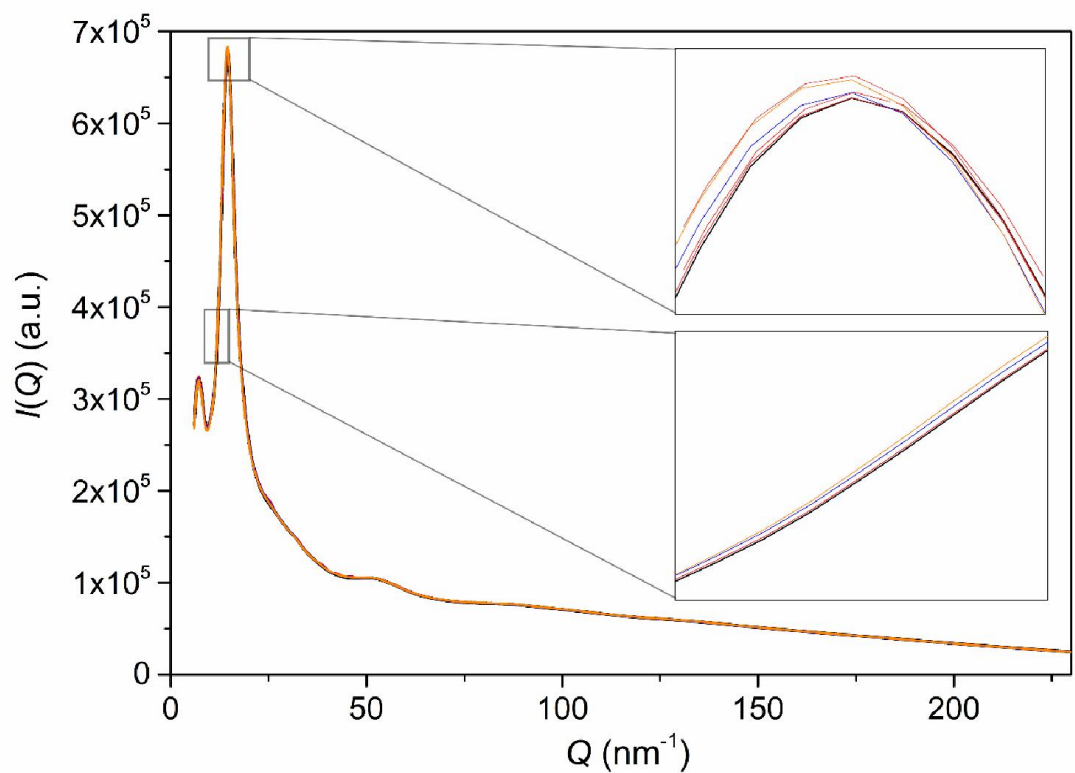

**Figure S1:** Comparison of $I(Q)$ for pure propanol with $I(Q)$ of the redispersed ZnO NPs in propanol. The ligand-capped ZnO NPs (acetate = red curve, citrate = blue curve, pent = orange curve) induce a slight peak shift and change of the FWHM of the main peak in comparison to pure propanol (black). See insets for magnification of most distinct features.

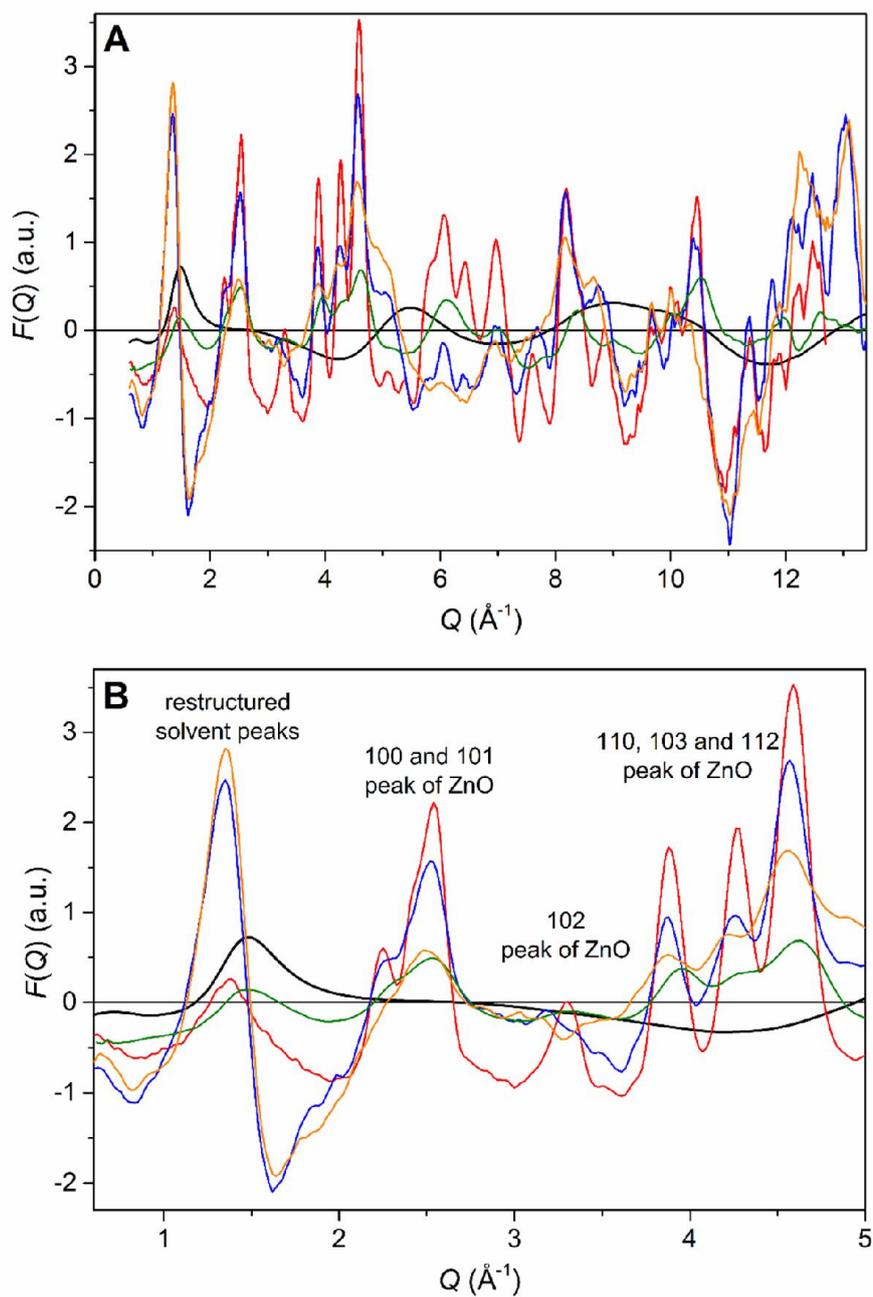

**Figure S2:** Change of reduced total scattering structure function $F(Q)$ by redispersion of NPs. The $F(Q)$ of pure propanol (black) features one distinct peak at 1.5 Å$^{-1}$ (see magnification in b for more detail) and a long-frequency oscillation describing the SRO. The $F(Q)$s of the ZnO NPs in propanol (acetate = red, citrate = blue, dmlt = green, pent = orange) contain sharp peaks from the NPs, but additionally a sharp peak from the restructured solvent between 1 and 2 Å.

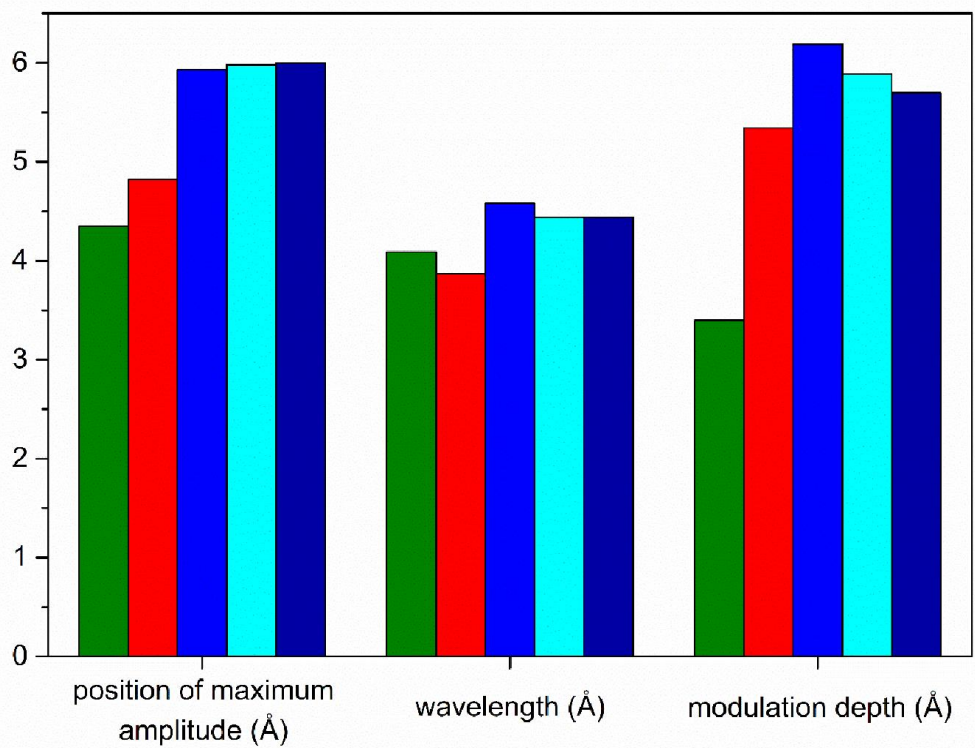

**Figure S3:** Influence of flow velocity on restructuring. Fit results of the position of the maximum amplitude, the wavelength and modulation depths of ZnO NPs with ligand pent redispersed in different solvents with different flow velocities: methanol at 190 mL/min (green), ethanol at 190 mL/min (red), propanol at 100 mL/min (light blue), propanol at 190 mL/min (purple), propanol at 300 mL/min (dark blue). The change of the parameters, i.e. the restructuring, for different pumping speeds in the same solvent is smaller than the difference in between different solvents at the same pumping speeds of 190 mL/min.

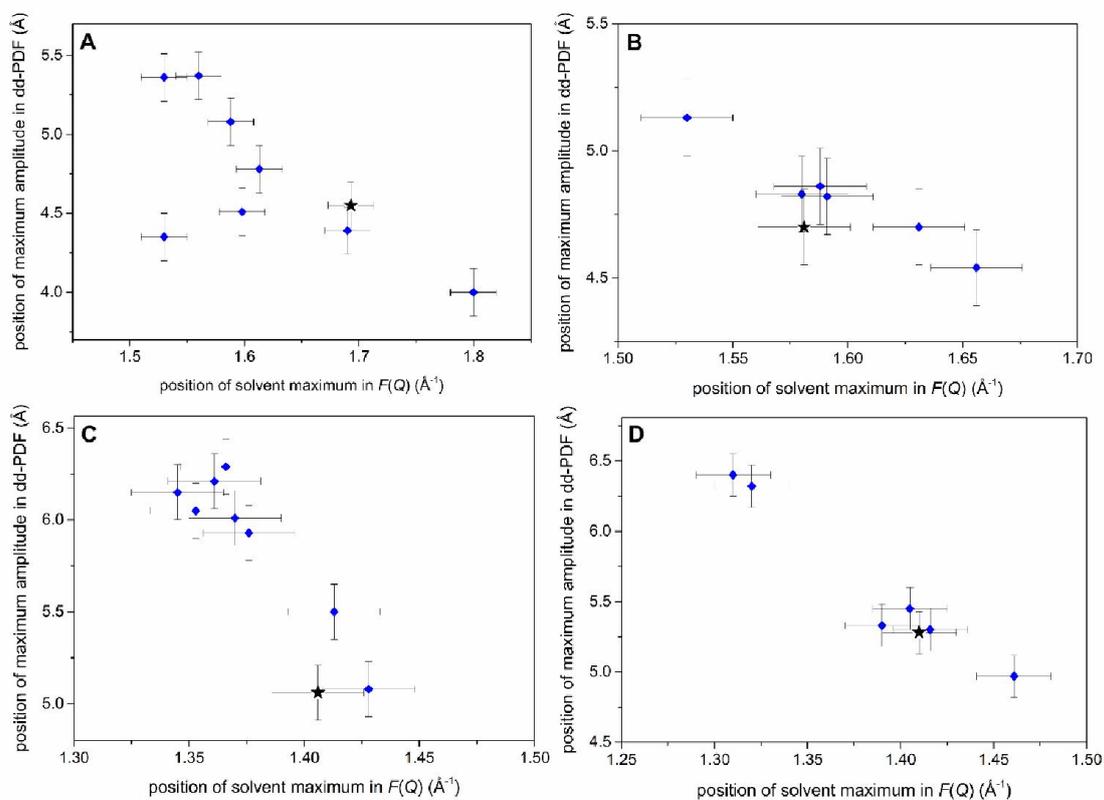

**Figure S4:** Correlation between the position of the first maximum in the dd-PDFs and the position of the solvent maximum in the $F(Q)$s. Fits of Pseudo-Voigt functions to the remainder of the restructured solvent in the $F(Q)$ data give the position of the solvent maximum, which is clearly correlated to the position of the maximum amplitude in the PDF. Black stars represent the parameters of the pure solvent, blue rhombs are from all NPs. Panel A shows methanol, B ethanol, C propanol and D hexane. Error bars resulted from least-squares fits.

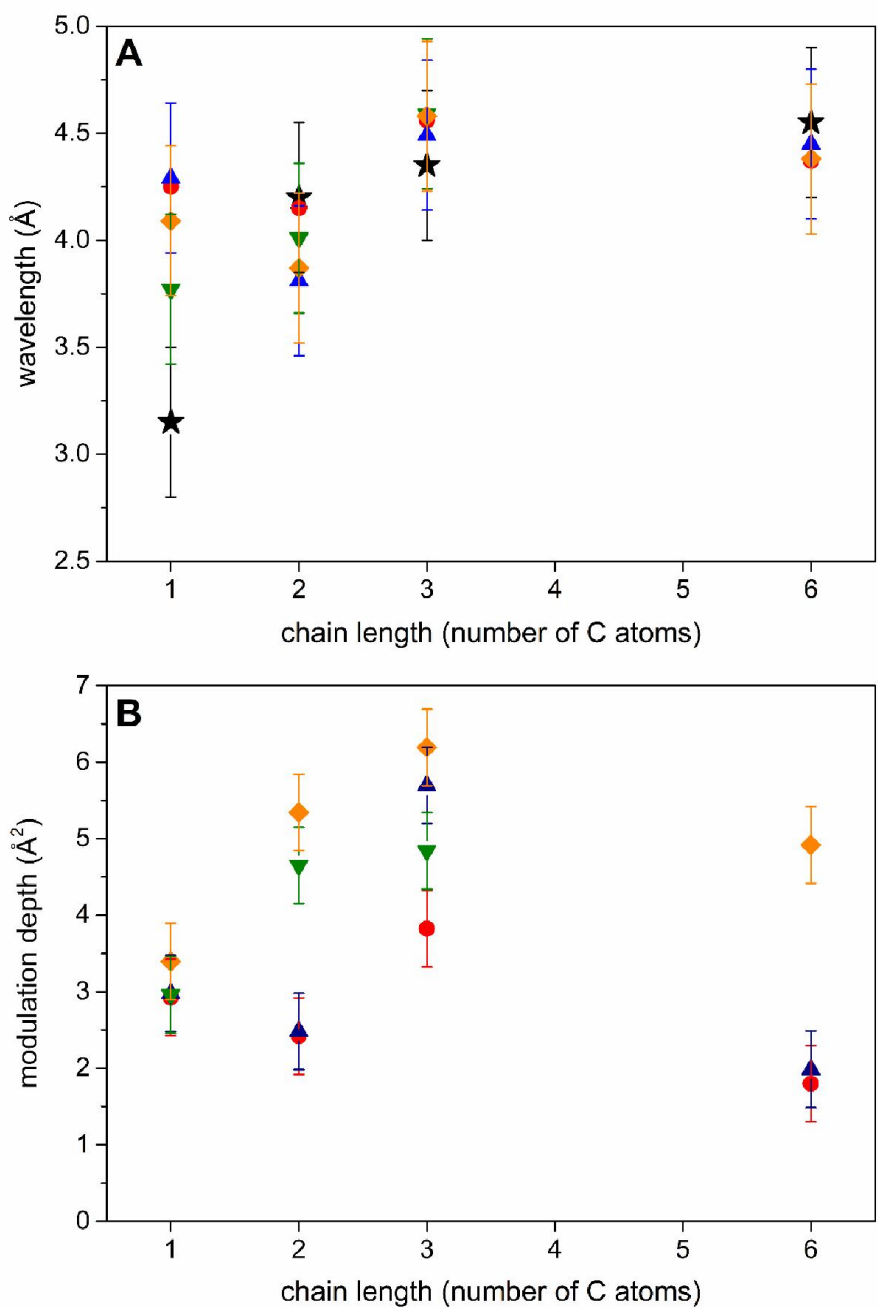

**Figure S5:** Parameters of restructured solvent for ZnO NPs. Wavelength (panel A) and modulation depth (panel B) plotted over the chain length of the organic solvent. Black stars represent the pure solvent, the other filled symbols are for the four different ZnO NPs capped with acetate (red circle), citrate (blue triangle), dmlt (green triangle), pent (orange rhombs). Error bars resulted from least-squares fits.

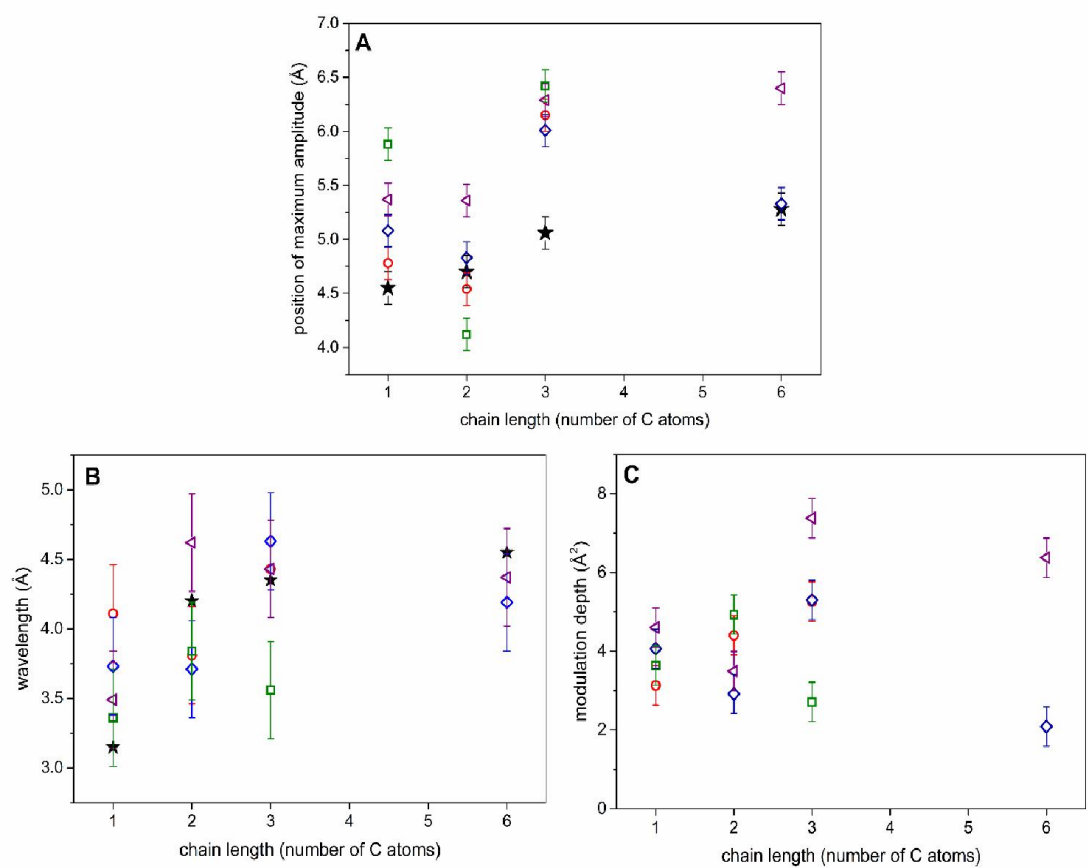

**Figure S6:** Parameters of restructured solvent for Ag, TiO$_2$, ZrO$_2$ as well as In$_2$O$_3$ NPs. Position of the maximum amplitude (panel A), wavelength (panel B) and modulation depth (panel C) plotted over the chain length of the organic solvent. Black stars represent the pure solvent, the other empty symbols are for the different NPs, i.e. silver (red circle), TiO$_2$ (blue rhomb), ZrO$_2$ (purple triangle)), In$_2$O$_3$ (green square). Error bars resulted from least-squares fits.

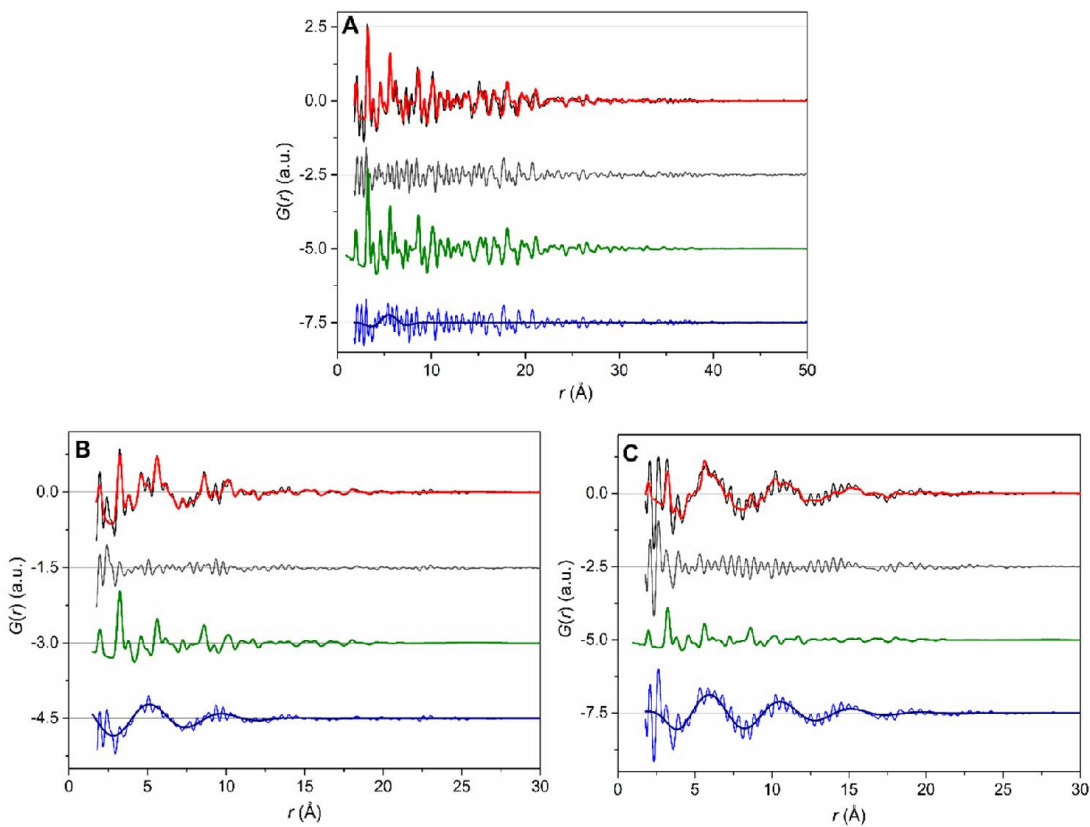

**Figure S7:** PDF fits of redispersed ZnO NPs with illustration of single contributions. Experimental d-PDF of ZnO NPs with different ligands (black) and their fit (red), showing the overall difference of the fit (grey), the contribution of the nanoparticle (green) and the contribution and fit of the restructured solvent (blue). The single contributions are shifted for clarity. Panel A displays the ZnO NPs with ligand acetate, panel B with dmlt and panel C with pent.

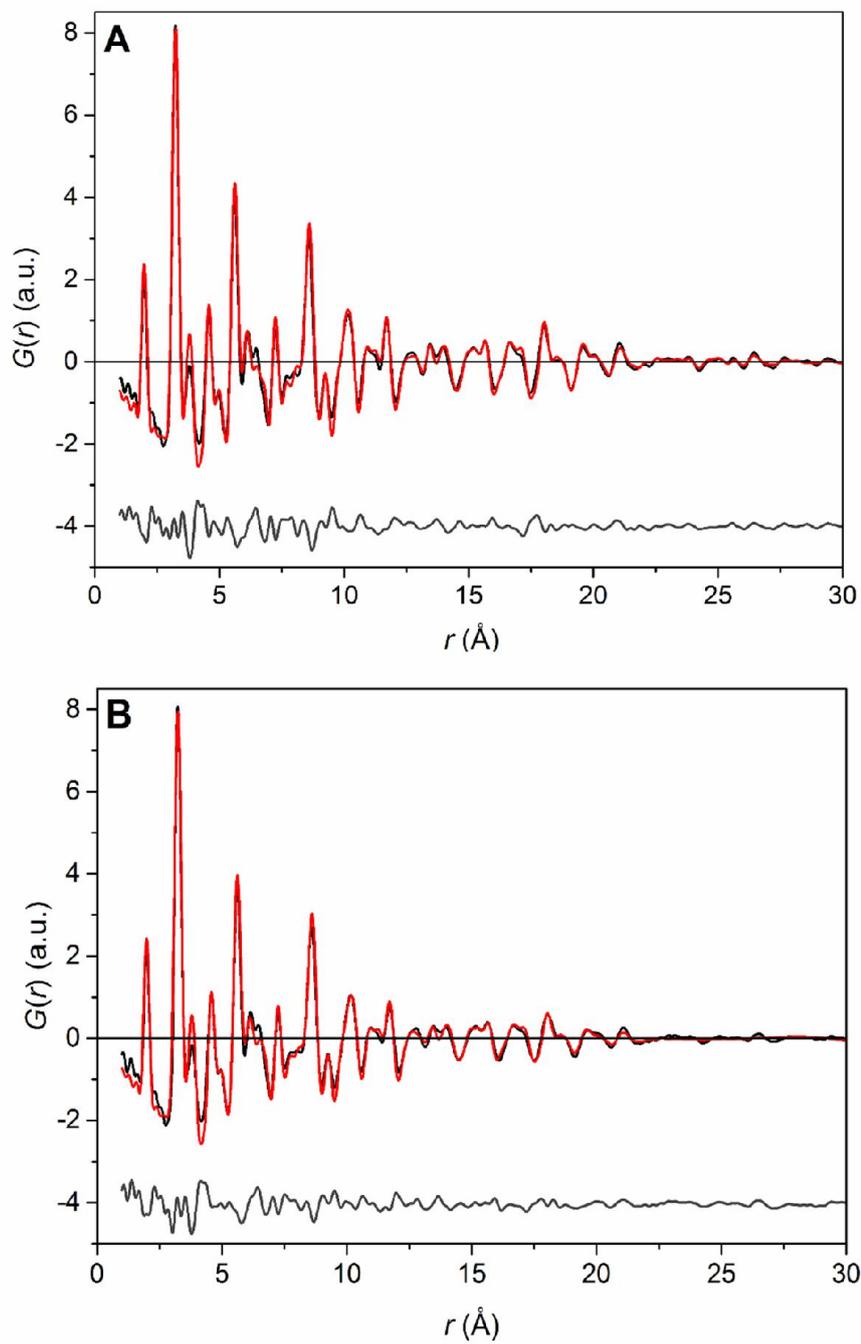

**Figure S8:** PDF fit of ZnO NP powder with ligand citrate (A) and pent (B). Experimental PDFs of ZnO NP powders (black) and their fits (red), with the difference curves in offset (black).

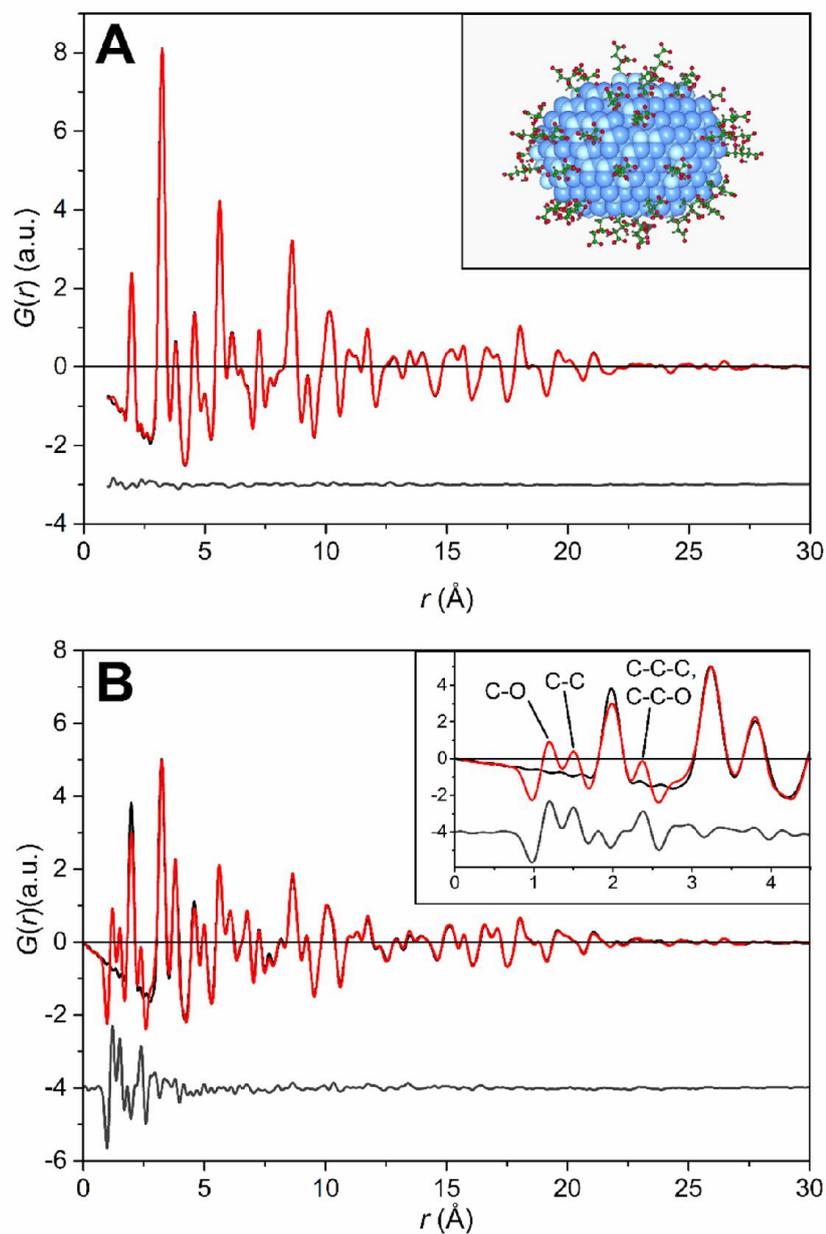

**Figure S9:** Modelled x-ray and neutron PDFs of ZnO NP powders with and without ligand citrate. For x-rays (panel A), the PDFs without ligands (black) and with ligands (red) are almost identical, resulting in a difference curve, neutron PDF minus x-ray PDF (black, in offset) which is almost a straight line. The inset shows the modelled ZnO NP with ligand citrate. For neutrons (panel B), the ligand molecules cause distinct sharp peaks below 5 Å, which can be attributed to distinct intramolecular distances (see inset), but do not create any sinusoidal oscillation, which could resemble solvent restructuring.

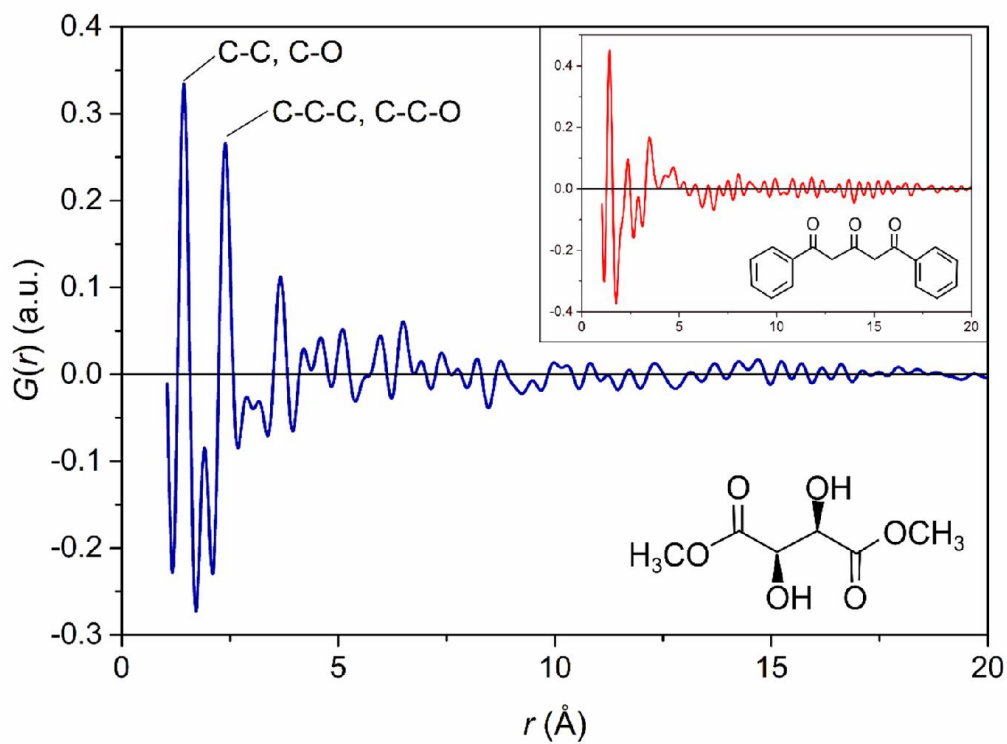

**Figure S10:** d-PDFs of dissolved ligand molecules in ethanol. The d-PDFs only show intramolecular distances of the dmlt (blue) and pent (red, inset) molecules, when these are dissolved at concentrations used in the synthesis of the ZnO NP. The structural formulas of the molecules are shown each at the bottom right corner.

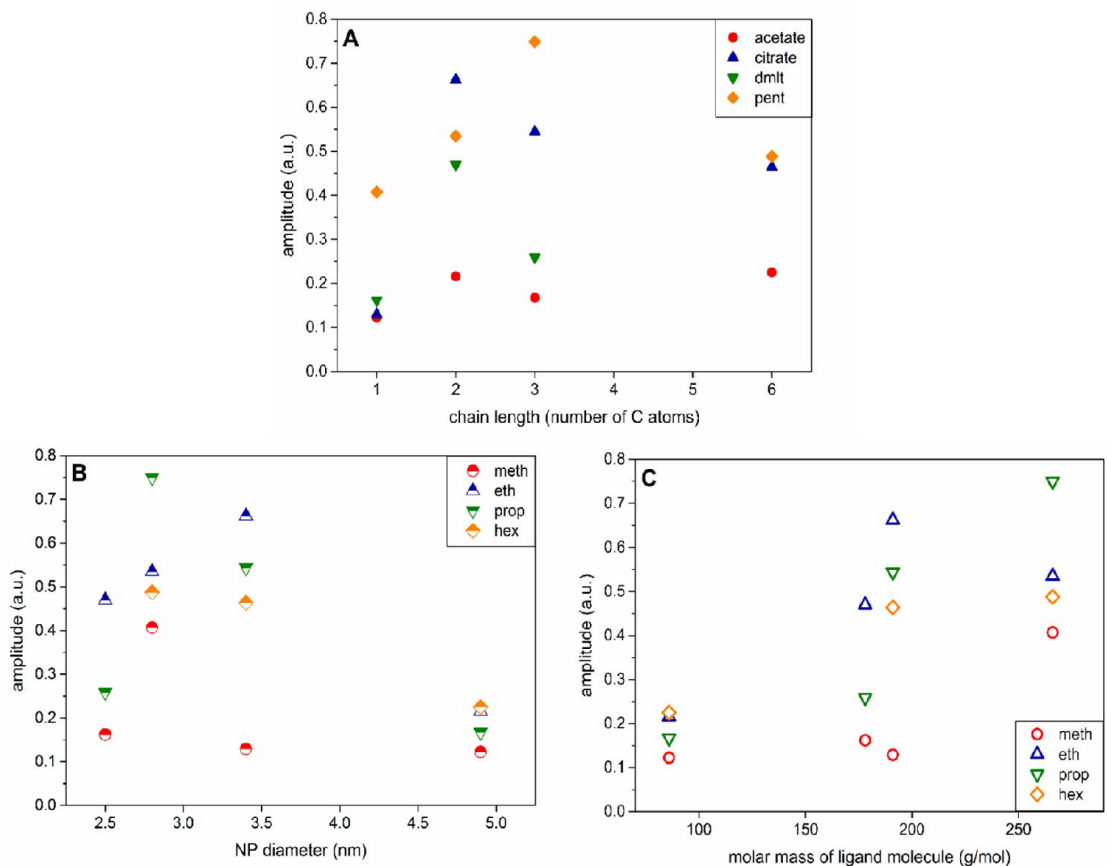

**Figure S11:** Dependence of amplitude of restructured solvent around ZnO NPs on different parameters including chain length of solvent (panel A), nanoparticle size (panel B) and molar mass of ligand molecule (panel C).

**Table S1:** PDF fit results of ZnO nanoparticle powders with different ligands.

| Ligand | Diameter ab-plane | Diameter along c-axis | Stacking fault probability | Lattice parameter ab | Lattice parameter c | B-value | z position of Zn atom |
|---|---|---|---|---|---|---|---|
| | Å | Å | a.u. | Å | Å | a.u. | a.u. |
| **acetate** | 49.2 | 49.3 | 0.11 | 3.2587 | 5.23916 | 0.716 | 0.3764 |
| **citrate** | 33.8 | 31.5 | 0.25 | 3.2458 | 5.23117 | 0.801 | 0.3806 |
| **dmlt** | 24.8 | 22.6 | 0.22 | 3.2451 | 5.23123 | 1.047 | 0.3724 |
| **pent** | 27.5 | 20.7 | 0.29 | 3.247 | 5.23047 | 0.929 | 0.3730 |